\magnification=\magstep1
\baselineskip=30pt
\input amssym.def
\input amssym.tex

\hskip5truein 6/23/03

\centerline{\bf On the (Boltzmann) Entropy of Nonequilibrium
Systems}
\centerline {{\bf
S. Goldstein}\footnote{*}{oldstein@math.rutgers.edu} and {\bf Joel L.
Lebowitz}\footnote{**}{lebowitz@sakharov.rutgers.edu}}
\centerline{\bf Department of Mathematics and Physics}
\centerline {\bf Rutgers University}
\centerline {\bf Piscataway, New Jersey}

\noindent {\bf Abstract:} Boltzmann  defined the entropy of a
macroscopic system in a macrostate $M$ as the $\log$ of the volume of
phase space (number of microstates) corresponding to $M$.  This agrees
with the thermodynamic entropy of Clausius when $M$ specifies the
locally conserved quantities of a system in local thermal equilibrium
(LTE).  Here we discuss Boltzmann's entropy, involving an appropriate
choice of macro-variables, for systems not in LTE.  We generalize the
formulas of Boltzmann for dilute gases and of Resibois for hard sphere
fluids and show that for macro-variables satisfying any deterministic
autonomous evolution equation arising from the microscopic dynamics
the corresponding Boltzmann entropy must satisfy an ${\cal
H}$-theorem.


\noindent PACS NUMBERS:  05.20.-y, 05.20.Dd, 05.20.Jj, 05.70.Ln

\bigskip \bigskip

\noindent {\bf Introduction}

Thermodynamics associates to isolated equilibrium macroscopic systems
with specified thermodynamic parameters $M$ an additive,
macroscopically well defined, entropy $S(M)$.  The second law of
thermodynamics then asserts that in any temporal change occurring in
such isolated systems (due e.g.\ to the relaxation of some constraint)
the new equilibrium state, with parameters $M^\prime$, must satisfy
$S(M^\prime) \geq S(M)$.  $S(M)$ as well as the second law naturally
generalize to an entropy $S_{\rm loc. eq.}(M) = S_{\rm
loc. eq.}(\{M({\bf x})\})$ for systems in LTE with particle, momentum
and energy densities varying slowly (on a microscopic scale) in space
and time.  $S_{\rm loc. eq.}(M)$ increases with time when $M({\bf x})$
evolves to $M_t({\bf x})$ according to macroscopic hydrodynamical
equations [1,2].  This is reviewed briefly in section 2.

We then discuss in section 3 Boltzmann's microscopic interpretation of
$S(M)$ as the $\log$ of the volume of phase space associated to $M$.
This not only provides a formula for computing $S(M)$ microscopically,
but also explains the origin of the time-asymmetric second law in the
time-symmetric microscopic laws [3, 4].  It shows in particular that
if there is a deterministic autonomous equation describing the time
evolution of a macrostate $M_t$ of an isolated system, be it
hydrodynamic or kinetic, e.g.\ the Boltzmann equation, it must give an
$S(M_t)$ which is monotone non-decreasing in $t$.

Boltzmann's macroscopic formulation leads naturally to a formula for
the entropy of dilute gases which may be far from LTE.  {}For such
systems the macrostate $M$ may be specified by $f({\bf x}, {\bf v})$,
the density of gas particles in the six-dimensional one-particle phase
space.  Boltzmann showed that this entropy, $S_{\rm gas}(f_t)$,
increases with time when $f_t$ evolves according to the Boltzmann
equation (${\cal H}$-theorem) [3, 4, 5].  This is discussed in section
4.

In section 5 we give a formula for $S(f, E)$, the log of the phase
space volume of a general system whose macrostate is specified by
$f({\bf x}, {\bf v})$ and the total energy $E$.  This reduces to
$S_{\rm gas}(f)$ for a dilute gas and to $S_{hs}(f)$ for a system of
hard spheres.  $S_{hs}(f)$ was found by Resibois [6] to satisfy an
$\cal H$-theorem when $f$ evolves according to the (modified) Enskog
equation for a system of hard spheres.  We note that the general
argument given by Boltzmann for the origin of the second law suggests
that $S(f_t,E)$ should be monotone in time even if $f_t$ does not
satisfy an autonomous evolution equation.  This is discussed further
in section 6.

Section 7 consists of some remarks comparing and contrasting
Boltzmann's definition of the entropy of a macroscopic system with
other definitions of entropy.  We raise, but do not resolve, the
question of the appropriate choice of macrostates for general
nonequilibrium systems.

The article is written in an informal style describing the ideas and
facts (not necessarily in the right historical order) we think
important for understanding the notion of the entropy of a macroscopic
system, made up of a very large number of atoms or molecules.  We
restrict ourselves to isolated classical systems, assume familiarity with the
basic notions of thermodynamics and statistical mechanics, and omit
many details (including units, boundary conditions, etc.).  We refer
the interested reader to [7] and references there.

\noindent {\bf 2.  Clausius' Macroscopic Entropy}

Rudolf Clausius' 1865 paper [8] concludes with his celebrated ``two
fundamental theorems of the mechanical theory of heat'': 1.  The
energy of the universe is constant.  2.  The entropy of the universe
tends to a maximum.  These express in succinct form what is generally
referred to as the first and second law of thermodynamics; see [3, 4]
for their interesting history.

The first law needs no elaboration.  The existence of a conserved
energy for isolated systems goes back to Newton for mechanical
systems.  The experiments of Joule then showed, that thermal phenomena
are subject to the same mechanical laws.

The second law, on the other hand, which contains the newly coined
word entropy, does need elaboration.  Let us quote Lars Onsager [9]:
``The second law of thermodynamics forbids perpetual motion of the
second kind and implies the existence of a definable entropy for any
system in a state that can be reached by a succession of reversible
processes.  These ``thermodynamic'' states are typically defined as
states of ``equilibrium'' under specified restraints on composition,
energy, and external boundary conditions, in the sense that no
spontaneous change can occur in the system as long as the constraints
remain fixed.''  The implicit ``restraints'' exclude chemical or
nuclear reactions which would change the species present, etc..

As put in the textbooks, e.g.\ [1]: given an equilibrium system with energy
$E$ and mole (or particle) numbers ${\bf N}$ in a spatial region $V$,
with a volume which we shall also denote by $V$, there exists a
function $S(E,{\bf N}, V)$ such that in a reversible process
$$
dS = [dE + pdV - \sum \mu_j dN_j]/T  \eqno(1)
$$
where $T$ is the absolute temperature, $p$ the pressure and $\mu_j$
the chemical potential of species $j$.  The terms in the square
bracket just give the amount of heat added to the system in a
reversible process.

Thermodynamics further states that the entropy of two isolated
macroscopic systems, each in equilibrium, with their own energies,
mole numbers and volumes, is the sum of their individual entropies,
i.e.
$$
S_{1,2}(E_1,{\bf N}_1,V_1.E_2,{\bf N}_2,V_2) = S_1(E_1,{\bf N}_1.V_1)
+ S_2(E_2,{\bf N}_2,V_2)\eqno(2)
$$

Suppose now that these two systems are permitted to interact and
exchange energy over some period of time after which they are again
isolated.  Then, according to the first law, their new energies
$E^{\prime}_1$ and $E^{\prime}_2$ will satisfy $E^{\prime}_1 +
E^{\prime}_2 = E_1 + E_2$.  If we now wait until each of the systems
comes to equilibrium then, according to the second law, their
combined new entropy must satisfy the inequality,
$$
S^\prime_{1,2} = S_1(E^\prime_1, {\bf N}_1, V_1) + S_2(E^\prime_2,
{\bf N}_2, V_2) \geq S_{1,2} = S_1(E_1,{\bf N}_1,V_1) + S_2(E_2,
{\bf N}_2, V_2).\eqno(3)
$$ 

Similar inequalities hold when relaxing other constraints.  As an
extreme example imagine that initially system 1 contained a mixture of
hydrogen and oxygen at a low temperature $T_1$, with an implicit
constraint prohibiting their chemical reaction, while system 2 was at
a high temperature $T_2$. The final state could now be very hot steam
in system 1, with a temperature $T_1^\prime$ higher than the
temperature $T_2^\prime$ in system 2, e.g.\ $T^\prime_1 > T_2^\prime >
T_2 > T_1$.

In all cases, {\it if the two systems have come to a joint
equilibrium} at the end of this period of interaction, with possibly
new ${\bf N}^\prime$, then $E^\prime_1$ and $E^\prime_2$ must be such
that the new entropy satisfies
$$
S^\prime_1(E^\prime_1,{\bf N}^\prime_1, V_1) +
S^\prime_2(E^\prime_2,{\bf N}^\prime_2, V_2) = S^\prime_{1,2} =
\sup_{U_1,U_2} \big\{S_1(U_1,{\bf N}^\prime_1,V_1) + S_2(U_2, {\bf
N}^\prime_2, V_2)\big \} \eqno(4)
$$
subject only to energy conservation, $U_1+U_2 = E_1+E_2$.  This
implies 

$$
T^\prime_1 \equiv \big [ {\partial S_1 \over \partial E^\prime_1}
(E^\prime_1, {\bf N}^\prime_1, V_1)\big ]^{-1} = T^\prime_2 \equiv \big [
{\partial S_2(E^\prime_2, {\bf N}^\prime_2, V_2) \over \partial E^\prime_2}
\big ] \eqno(5)
$$
Similar relations hold when there can be an exchange of matter or
volume between the systems.

Using the fact that for macroscopic systems surface areas (multiplied
by a suitable microscopic length) and surface energies are negligible
compared to the corresponding bulk quantities, the additivity of the
entropy, expressed by (2), also gives extensivity.  That is, for
systems uniform in the bulk,
$$
S(E, {\bf N}, V) =  V s(e,{\bf n}).\eqno(6)
$$
where $e = E/V$ and ${\bf n} = {\bf N}/V$.

The thermodynamic entropy can now be extended to a system in LTE [1, 2]:
a system in a volume $V$ which can be
considered, to a good approximation, as being locally in
equilibrium with energy density $e(\bf x)$, particle density $n(\bf
x)$ and hydrodynamic velocity ${\bf u}({\bf x})$: for a precise
definition see [10].  {}For such systems we can, by extension of (6),
write 
$$
S_{\rm loc. eq.}(n,{\bf u},e) = \int_V s(e({\bf x}) - {1 \over 2} mn({\bf x})
{\bf u}^2({\bf x}), n({\bf x}))d{\bf x}\eqno(7)
$$
where $m$ is the mass (per mole) and we consider just one component for
simplicity, [1--12].  (The ``mechanical'' energy associated with ${\bf u}({\bf
x})$ does not contribute to  $S$ until it is dissipated and we always take
$\int_V n({\bf x}) u({\bf x})d{\bf x} = 0$).

Eq.\ (7) clearly agrees with (6) when the system is in true
equilibrium, ${\bf u} = 0$, and $e$ and $n$ are independent of $\bf
x$.  {}Furthermore if a LTE state evolves in time according to
macroscopic equations then $S_{\rm loc. eq.}$ must increase (or at
least not decrease) as a function of $t$.  Consider for example an
isolated system in LTE (with ${\bf u}=0$ and $n$ constant) with an
energy density profile $e_0({\bf x})$ and corresponding temperature
profile $T_0({\bf x})$ for which we have, using an extension of (1) to
continuous time,
$$
{\partial s \over \partial t} = -{\nabla \cdot {\bf J} \over T} =
-\nabla \cdot({\bf
J}/T) + {\bf J} \cdot \nabla(1/T), \eqno(8)
$$
where $\bf J$ is the heat flux, ${\bf J}/T$ is the entropy flux, and
${\bf J}\cdot\nabla(1/T)$ is the entropy production.  According to
{}Fourier's law, ${\bf J} = -{\cal K} (T) {\bf \nabla}T$ so (8) can be
written in the more familiar form as
$$
C_V(T) {\partial \over \partial t} T({\bf x},t) = 
\nabla \cdot [{\cal K}(T) \nabla T], \eqno(9)
$$
where $C_V(T) = T^{-1} {ds(T) \over dT}$ is the specific heat and
$\cal K$ is the heat conductivity.  Eq.\ (9) is to be solved, for $t >
0$, subject to no heat flux, or $\nabla T = 0$, at the surface of $V$.
Integrating (8) or (9) then yields
$$
{dS_{\rm loc. eq.} \over dt} = {d \over dt} \int_V s d{\bf x} = \int_V
{\bf J} \cdot (\nabla {1 \over T})d{\bf x} = \int_V {\cal K} T^2
(\nabla {1 \over T})^2 d{\bf x} \geq 0.
\eqno(10)
$$
As $t \to \infty$, $T({\bf x},t) \to \bar T$, for all ${\bf x} \in V$,
with $\bar T$ determined by energy conservation and $S_{\rm loc. eq.}$
approaches its maximum (equilibrium) value $Vs({\bar e}, n)$.

The second law thus manifests itself for LTE by the requirement that
the entropy production ${\bf J} \cdot \nabla(1/T)$ be non-negative,
i.e.\ that ${\cal K}(T) \geq 0$. The formula for entropy production
generalizes to other macroscopic equations [2, 12], e.g.\ to the
compressible Navier-Stokes equations which describe the time evolution
of $n({\bf x},t), {\bf u}({\bf x},t)$ and $e({\bf x},t)$.  The Euler
equations, on the other hand, conserve $S_{\rm loc. eq.}$, in the
absence of singularities of the flow.  They do not give an approach to
equilibrium of an isolated system and thus do not provide a
description valid for times over which such an approach takes place.
In the presence of shocks, however, the solutions are non-unique and
the requirement that $S_{\rm loc. eq.}$ increase picks out the correct
evolution.

\noindent {\bf 3.  Boltzmann's Microscopic Entropy}

There are many ways to stretch the notion of LTE and apply the second
law to processes taking place in systems which are clearly very far
from equilibrium, e.g.\ living organisms [3, 9].  These ad hoc
extensions work quite well in the hands of seasoned practitioners [9]
but are far from systematic.  It would certainly be desirable to find
systematic ways for defining and calculating the entropy, expressed as
a function of the appropriate macroscopic variables of systems that
are not in LTE.  This entropy would be monotone in time and coincide
with $S_{\rm loc. eq.}$ for a system in LTE.

This is exactly what was accomplished by Boltzmann's microscopic
interpretation of the macroscopic Clausius equilibrium entropy $S(M)$.
This interpretation provides a formula for the computation of $S(M)$
from the microscopic Hamiltonian.  Even more importantly, it explains
the origin of the time-asymmetric second law in the time-reversible
dynamics of the atoms and molecules which are the microscopic
constituents of macroscopic matter, and shows its applicability to
systems not in LTE.  We will be very brief here, c.f.\ [1--14].

A very good summary of Boltzmann's accomplishment is given by this
quote from Einstein [13]: ``On the basis of the kinetic theory of
gases Boltzmann had discovered that, aside from a constant factor,
entropy is equivalent to the logarithm of the ``probability'' of the
[macro]state under consideration.  Through this insight he recognized
the nature of the course of events which, in the sense of
thermodynamics, are ``irreversible''.  Seen from the
molecular-mechanical point of view, however, all courses of events are
reversible.  If one calls a molecular-theoretically defined state a
microscopically described one, or, more briefly, micro-state, then an
immensely large number $(Z)$ of states belong to a macroscopic
condition. $Z$ then is a measure of the probability of a chosen
macro-state.  This idea appears to be of outstanding importance also
because of the fact that its usefulness is not limited to microscopic
description on the basis of mechanics.''

Let us make Einstein's remarks more explicit by considering a
classical system of $N$ particles in a box $V$.  Its microstate $X$ is
given by a point in the $6N$ dimensional phase space which specifies
everything about the system, e.g.\ the energy given by its Hamiltonian
$H(X)$, etc..  When $N$ is very large a more appropriate
coarse-grained description of the system is provided by its macrostate
$M$.  We can specify $M$, for example, by dividing $V$ into $J$ cubes
$\Delta_k$, $J << N$, so that each cube contains a very large number
of particles and specifying coarse-grained values of the energy,
momentum and number of particles in each $\Delta_k$. Let $\Gamma_M$ be
the region of the phase space consisting of all microstates consistent
with $M$, i.e.\ the set of all $X$ such that the appropriate phase
space function $M(X) = M$.  Let $|\Gamma_M|$ be the volume of
$\Gamma_M$ in appropriate units (this is Einstein's $Z$ for a
classical system).  Boltzmann defined the entropy of a macroscopic
system with microstate $X$ by
$$
S_B(X) = k \log|\Gamma_{M(X)}|.\eqno(11)
$$

Boltzmann then showed for a gas in an equilibrium macrostate,
$M_{eq}$, which, for the above choice of $M$, corresponds to a uniform
density of the macro-variables in $V$, that $S_B$ agrees (to leading order
in $N$) with the thermodynamic entropy of Clausius.  The same is true
for LTE states, i.e.\ if $M(X) = \{n({\bf x}), {\bf u}({\bf x}),
e({\bf x})\}$ then $S_B(M) = k \log |\Gamma_{M(X)}| = S_{\rm
loc. eq,}(n, {\bf u}, e)$.  This means that if the entropies $S_B(M)
> S_B(M^\prime)$ differ by a macroscopic amount, the ratio of their
corresponding phase space volumes is exponentially large in $N$.  Thus
if the system contains one mole of material, the ratio of
$|\Gamma_{M_{eq}}|$ to $|\Gamma_M|$ for a macrostate $M$ in which all
the particles are in the left half of the box is of order
$\exp[10^{20}]$.  This is far larger than the ratio of the volume of
the known universe to the volume of one proton.

Boltzmann then argued that given this disparity in sizes of $\Gamma_M$
for different $M$'s, the time evolved $M_t = M(X_t)$ will be such that
$|\Gamma_M(X_t)|$ and thus $S_B(X_t)$ will {\it typically} increase in
accord with the second law.  By ``typically'' we mean that for any
$\Gamma_M$ (of the kind described above) the relative volume of the
set of microstates $X$ in $\Gamma_M$ for which the second law is
violated by a macroscopic amount, i.e.\ by an amount proportional to
$N$, during any fixed time period (not bigger than the age of the
universe), goes to zero rapidly (exponentially) in the number of atoms
and molecules in the system.

In fact let us consider the case where $M_t$ satisfies an autonomous
deterministic evolution, e.g. equation (9).  This means that if that
evolution carries $M_{t_1} 
\to M_{t_2}$, then the microscopic dynamics $\Phi_t$ carries
$\Gamma_{M_{t_1}}$ inside $\Gamma_{M{t_2}}$, i.e. $\phi_{t_2-t_1}
\Gamma_{M_{t_1}} \subset \Gamma_{M_{t_2}}$, with negligible error.  
Now the fact that phase space volume is conserved by the Hamiltonian
time evolution (Liouville's theorem) implies that $|\Gamma_{M_{t_1}}|
\leq |\Gamma_{M_{t_2}}|$ and thus by (11) that $S_B(M_{t_2}) \geq
S_B(M_{t_1})$ for $t_2 \geq t_1$.  We have thus derived an ``${\cal
H}$-theorem'' for any deterministic evolution of the macro-variables
arising from the microscopic dynamics.  The explicit form for the rate
of change of $S_B(M_t)$ (including strict positivity) depends on the
detailed macroscopic evolution equation.  The fact that
$\Gamma_{M_{eq}}$ essentially coincides for large $N$ with the whole
energy surface $H(X) = E$ also {\it explains} the evolution towards
and persistence of equilibrium in an isolated macroscopic system.

The emergence of definite time-asymmetric behavior in the observed
evolution of macroscopic systems, despite the total absence of such
asymmetry in the microscopic dynamics, is thus a consequence of the
great disparity between microscopic and macroscopic scales, together
with the fact (or very reasonable assumption) that what we observe in
nature is typical behavior, corresponding to typical initial
conditions, c.f.\ [7].

\noindent {\bf 4.  Going Beyond LTE:  Dilute Gases}

As is clear from Eq.\ (11), the choice of the macro-variables $M$ is
essential for the computation of $S_B(X)$.  {}For equilibrium systems or
those in LTE these are specified by thermodynamics although there is a
large leeway in choosing the sizes of the boxes $\Delta_k$, if we
consider only leading terms in $N$. They are the locally conserved and
hence microscopically slowly varying quantities---precisely those for
which one has hydrodynamic type autonomous equations.

To obtain useful quantitative information from the second law for
systems not in LTE one has to find appropriate macro-variables $M$ for
the system under consideration, e.g.\ those which satisfy autonomous
time evolution equations, and for which one can compute $S_B(M)$.  A
paradigmatic example of where this has been achieved is a dilute gas.
{}Following Boltzmann, we refine the $M$ considered in the last section
for a system of $N$ particles in a box $V$.  This is done by noting
that the microstate $X =
\{{\bf r}_i, {\bf v}_i\}$, $i = 1,...,N$, can be considered as a set of
$N$ points in six dimensional $\mu$-space, somewhat analogous to
positions $\{{\bf r}_i\}$ in $V \subset {\Bbb R}^3$.  We may then divide up 
this $\mu$-space space into $\tilde J$ cells
$\tilde \Delta_\alpha$, centered on $({\bf r}_\alpha, {\bf v}_\alpha)$, of
volume $|\tilde \Delta_\alpha|$.  A macrostate $\tilde M$ is then
specified by the (coarse grained) number of particles in each $\tilde
\Delta_\alpha$, 
$$
\tilde M = \{N_\alpha\}, \quad \alpha = 1,...,\tilde J << N.
 \eqno(12)
$$

{}For dilute gases one can {\it neglect}, for typical configurations,
the existence of interactions between the particles, although of
course they still play a role in the dynamics now described by a
succession of collisions between pairs of particles [3, 4, 5].  Under
these conditions the coarse grained energy of the system in the state
$\tilde M$ is given by
$$ {1 \over 2} m \sum_\alpha N_\alpha {\bf
v}^2_\alpha = E \eqno(13)
$$
$$
\sum N_\alpha = N \eqno(14)
$$
We do not therefore need to specify the energy separately and 
the phase space volume associated with such an $\tilde M$ 
is then readily computed to be [4b] 
$$
| \Gamma_{\tilde M}| = \Pi_\alpha(N_\alpha!)^{-1}
|\tilde \Delta_\alpha|^{N_\alpha}  \eqno(15)
$$
where we do not distinguish between configurations in which particle
labels are interchanged.  {}For large enough $N$ and a judicious choice
of the $\{\tilde \Delta_\alpha\}$ we can, for almost all $X$
consistent with (13) and (14), use Stirling's formula in (15) 
and obtain
$$
S_B(\tilde M) \sim  -k\{\sum_\alpha ({N_\alpha \over
|\tilde \Delta_\alpha|} \log {N_\alpha \over |\tilde \Delta_\alpha|})
|\tilde \Delta_\alpha|- N\}. \eqno(16)
$$

Using $\tilde M$ we can associate with a typical $X$ a coarse grained
density $f_X \sim N_\alpha/|\tilde \Delta_\alpha|$ in $\mu$-space,
i.e.\ such that $N_\alpha =
\int_{\tilde \Delta_\alpha} d{\bf x} d{\bf v} f_X({\bf x},{\bf v})$. 
Eq.\ (16) then shows that, up to a constant (depending on $N$), the
Boltzmann entropy $S_B(X)$ is given by 
the negative of Boltzmann's ${\cal H}$-function, 
$$
S_{\rm gas} (f) = -k \int_V d{\bf x} \int_{{\Bbb R}^3} d{\bf
v} f({\bf x},{\bf v}) \log f({\bf x},{\bf v})   \eqno(17)
$$
where $f = f_X$.  (We shall drop the subscript $X$ unless we want to
emphasize that $f$ is associated with a given microstate $X$.)  The
maximum of $S_{\rm gas}(f)$ over all $f$ which satisfy the conditions,
$$
\int_V d{\bf x} \int_{{\Bbb R}^3} d{\bf v} f({\bf x}, {\bf v}) = N \eqno(18)
$$
$$
\int_V d{\bf x} \int_{{\Bbb R}^3} d{\bf v} {1 \over 2} m{\bf v}^2
f({\bf x}, {\bf v}) = E \eqno(19)
$$
is given by the equilibrium distribution
$$
f_{eq} = {N \over V}(2\pi k T/m)^{-3/2} \exp[- m{\bf v}^2/2kT] \eqno(20)
$$
where $kT = 2/3 (E/N)$.  $f_{eq}$ coincides of course with the density
$f_X({\bf x}, {\bf v})$ obtained for a typical microstate $X$ on the
energy surface $H(X) = E$ when $N$ is macroscopic (with deviations
going to zero as $N \to \infty$).

When $f \ne f_{eq}$ then $f$ and consequently $S_{\rm gas}(f)$ will
change in time.  The microscopic version of the second law, discussed
in section 3, now says that typical $X \in
\Gamma_{\tilde M}$ at the initial time $t=0$, will 
have an $\tilde M_t = \tilde M(X_t)$ with the property that
$S_B(\tilde M(X_t)) \geq S_B(\tilde M(X_{t^\prime}))$, for $t \geq
t^\prime$.  This means that $f_{X_t}({\bf x}, {\bf v}) = f_t({\bf x},
{\bf v})$ has to be such that $S_{\rm gas}(f_t) \geq S_{\rm
gas}(f_{t^\prime})$, for $t \geq t^\prime$.  This is exactly what
happens for a dilute gas for which the time evolution of $f_t({\bf
x},{\bf v})$ is well described by the Boltzmann equation (BE) [3--5]
which we shall not write out here: see [14] for a rigorous derivation
of the BE under suitable conditions.  As shown by Boltzmann, in his
famous ${\cal H}$-theorem, it indeed follows from the BE that ${d
\over dt} S_{\rm gas} (f_t)
\geq 0$, with equality holding only
when $f({\bf x}, {\bf v})$ is a local Maxwellian 
$\hat f({\bf v};n, {\bf u},T)$, [5]
$$
\hat f = n({\bf x})[{2\pi k T({\bf x}) \over m}]^{-3/2} \exp
\big \{ {-m[{\bf v} - {\bf u}({\bf x})]^2 \over 2 k T({\bf x})}
\big \}, \eqno(21)
$$

Eq.\ (21) defines LTE for a dilute gas 
with the $n({\bf x}), {\bf u}({\bf x})$ and $e({\bf x})$ obtained from
$f$ in the usual way, $n = \int_{{\Bbb R}^3} fd{\bf v}, {\bf u}({\bf
x}) = \int {\bf v} fd{\bf v}/n$, 
$kT({\bf x}) = {2 \over 3}
[e({\bf x}) - {1 \over 2} m n({\bf x}) {\bf u}^2 (x)]/n({\bf x})$.
When (21) is substituted into (17) we obtain
$$
S_{\rm gas}(\hat f) = \int_V d{\bf x} s_{\rm gas} (e,n) \eqno(22)
$$
with 
$$
s_{\rm gas}(e,n) = k \big \{ {3 \over 2} n \log (kT) - n(\log n -1)
\big \} + {\rm Const.}, \eqno(23)
$$
the Clausius entropy density given in (7), for a gas in LTE.  Since
$\hat f$ is not stationary unless $n, e$ and $\bf u$ are uniform in
the whole box, i.e.\ $\hat f = f_{eq}$, it is expected and partially
proven [5, 15], that starting with an initial $f_0({\bf x}, {\bf v})$,
which can be far from a local Maxwellian, $f_t({\bf x}, {\bf v})$ will
``rapidly'' approach an $f$ which is close to $\hat f({\bf v}; n, {\bf
u}, T)$ and will stay close to it while the local variables $n, {\bf
u}$ and $e$ change on a slower time scale.  As the gradients become
smaller this evolution will be hydrodynamic, i.e.\ $n, {\bf u}, e$
will evolve according to the compressible Navier-Stokes equations,
which will then bring the gas to equilibrium with $S_{\rm loc. eq.}$
increasing with time.

Note that $f$ satisfies the requirements for macro-variables discussed
in the beginning of this section so that $S_{\rm gas}(f)$ is indeed a
useful entropy functional.  The non-decrease of $S_{\rm gas}(f_t)$ for
$f_t$ a solution for the BE is, as already noted, a consequence of
Boltzmann's interpretation of the second law.  As put by Boltzmann:
``In one respect we have even generalized the entropy principle here,
in that we have been able to define the entropy in a gas that is not
in a stationary state'' [4b, p.\ 75].

It is important to distinguish between $f_{X_t}({\bf x}, {\bf v})$ and
another object with the same name, the marginal one-particle
(probability) distribution $F_1({\bf x}, {\bf v}, t)$ obtained from an
$N$-particle ensemble density evolving according to the Liouville
equation.  We mention here an instructive example in which $F_1({\bf
x}, {\bf v},0) = f_{X_0}({\bf x}, {\bf v})$ but $F_1({\bf x}, {\bf
v},t) \ne f_{X_t}({\bf x}, {\bf v})$ so that $F_1({\bf x}, {\bf v},t)$
does not give an adequate description of the macrostate of the system.
Consider a macroscopic system of $N$ noninteracting point particles,
moving among a periodic array of scatterers in a macroscopic volume
$V$, [7,10].  Starting with a nonuniform initial density
$f_{X_0}({\bf x}, {\bf v})$ the time evolved $f_{X_t}({\bf x}, {\bf
v})$ will approach an $f$ which depends only on $|{\bf v}|$ and which
will have a larger $S_{\rm gas}(f)$.  Since, however, $F_1({\bf x},
{\bf v},t)$ evolves according to the one-particle Liouville equation, $
\int\int F_1 \log F_1 d{\bf x} d{\bf v}$ remains constant in time.
What is crucial here is that what one might have regarded as the
obvious evolution equation for $f_{X_t}$ for this system, namely the
one-particle Liouville equation, in fact does not describe the
evolution of $f_{X_t}$ for times after which $F_1({\bf x}, {\bf v},
t)$ has developed structure on the microscopic scale.
(We note that when the periodicity of the scatterers is on the
microscopic scale then for macroscopic times the spatial density
profile $n_{X_t}({\bf x})$ will satisfy a diffusion equation [7, 10].)

\noindent {\bf 5.  The Boltzmann Entropy of Dense Fluids Not in LTE}

As already noted, it is very important for the microscopic derivation
of the second law and ipso facto for the increase of $S_{\rm
gas}(f_{X_t})$ that the initial microstate $X_0$ of the system under
consideration be typical of $\Gamma_{ M_0}$.  Consider now the case
when the interaction potential energy between the particles is not
negligible so that (19) is just the kinetic energy $K$ rather than the
total energy $E$.  The region $\Gamma_{\tilde M}$ will then include
phase points with widely differing total energies.  The set of
microstate $X$ of a system with a specified energy, $H(X) = E$ will
then correspond to a small fraction of $\Gamma_{{\tilde M}(X)}$ unless
$E$ is such that almost all of the points in $\Gamma_{{\tilde M}(X)}$
are also in the energy shell around $E$.

This illuminates the example considered by Jaynes [16]: One starts
with a nonequilibrium system with energy $E$ in which the kinetic
energy $K$ is larger than what it would be for an equilibrium system
with the same energy $E$.  The system will evolve towards equilibrium
with a Maxwellian velocity distribution at temperature $T_\infty = {2
\over 3k} K_E$ where $K_E$ is the equilibrium kinetic energy
corresponding to $E$.  Assuming that $f_0$ is a Maxwellian with
temperature $T_0$ we will have $T_\infty < T_0$ and so $S_{\rm
gas}(f_\infty) < S_{\rm gas}(f_0)$.  However, since the microstates
corresponding to the initial situation just described are not typical
for the macrostates $\Gamma_{f_0}$, this example does not contradict
the typical second law behavior of $S(f_t)$.

To properly describe non LTE macrostates for dense fluids let us
suppose now that both $f({\bf x}, {\bf v})$ and $E$ are used to
specify the macrostate $\tilde M$ for a fluid with Hamiltonian $H$
$$
H  = \sum {1 \over 2} m{\bf v}_i^2 + \sum \phi ({\bf r}_i - {\bf
r}_j)  \eqno(24)
$$
in which interactions are important.  We expect then that the
evolution of a typical initial $X_0 \in \Gamma_{M_0}$, with $H(X_0) =
E$ and $f_{X_0}({\bf x}, {\bf v}) = f_0({\bf x}, {\bf v})$, will
indeed be such that $S_B(\tilde M(X_t)) = S_B(f_t,E)$ will increase
with time in an actual system {\it even} if there is no autonomous
evolution law for $f_t$.

To compute $S_B(f,E) = \log
|\Gamma_{f,E}|$, 
let us consider first the case where the macrostate $\tilde M$ is specified
by both $f({\bf x}, {\bf v})$ and the local energy density
$e({\bf x})$.  The particle, kinetic and potential energy
densities  $n({\bf x})$, $K({\bf x})$ and
$\Phi({\bf x})$ are
determined from $f$ and $e$,
$$
n({\bf x}) = \int_{{\Bbb R}^3} f({\bf x}, {\bf v})d{\bf v}, 
\quad K({\bf x}) = {1 \over 2} m \int {\bf v}^2 f d{\bf v}, \quad
\Phi({\bf x}) = e({\bf x}) - K({\bf x}) \eqno(25)
$$
It is easy to see that the entropy corresponding to $M$ can be split
into momentum space and configuration space contributions
$$
S_B(f, e) = S^{(m)}(f) + S^{(c)}(n,\Phi)  \eqno(26)
$$
The momentum contribution $S^{(m)}$ can be readily computed along the
lines of formulas (15)--(17),
$$
S^{(m)}(f) = -\int_V d{\bf x} \int_{{\Bbb R}^3} d{\bf v} f({\bf x},
{\bf v}) \log[f({\bf x}, {\bf v})/n({\bf x})] \eqno(27)
$$
while $S^{(c)}(n,\Phi)$ is the configurational local equilibrium entropy
corresponding to the density $n({\bf x})$ and the potential energy
density $\Phi({\bf x})$.  $S^{(c)}$ is clearly the same as the
configurational part of $S_{\rm loc. eq.}$ computed at the energy
density $e^\prime({\bf x})$ that corresponds to $\Phi({\bf x})$ in
LTE, i.e.
$$
S^{(c)}(n, \Phi) = S_{\rm loc. eq.}(n, 0, e^\prime) - {3 \over
2} k \int_V n({\bf x}) \log T^\prime({\bf x}) d{\bf x} \eqno(28)
$$
where $T^\prime({\bf x})$ is the temperature corresponding
to $e^\prime({\bf x})$.
$S_{\rm loc. eq.}$ is defined in (7) using the equilibrium $s(e,n)$
and the subtracted kinetic term corresponds to the first term on the
right side of (23).

We can now obtain $S_B(f,E)$ by taking the $\sup$ of
$S_B(f, e)$ over all energy densities $e({\bf x})$ such 
that $\int_V e({\bf x}) d{\bf x} = E$.

An alternative way to 
compute $S_B(f, E)$ is to note that 
$$
S_B(f,E) = S^{(m)}(f) +
S^{(c)}(n,\Phi_{\rm tot.}) \eqno(29)
$$
where $S^{(c)}(n, \Phi_{\rm tot.})$ is the (configurational) entropy
associated with the macro-variables $n$ and $\Phi_{\rm tot.}$ for
$$
\Phi_{\rm tot.} = E - \int K({\bf x}) d{\bf x}. \eqno(30)
$$
We now observe that $S^{(c)}(n, \Phi_{\rm tot.})$ must agree with 
the configurational entropy of a system
with a Hamiltonian
$$
H^\prime = {m \over 2} \sum {\bf v}^2_i + \sum \phi({\bf r}_i - {\bf
r}_j) + \sum u({\bf r}_i) \eqno(31)
$$
in equilibrium at energy $E^\prime$, {\it when}
we choose $E^\prime$ and $u({\bf x})$ in such a way
that the equilibrium density is equal to $n({\bf x})$ and the total
(coarse grained) internal
potential energy has the value $\Phi_{\rm tot.}$, i.e.
$
\sum \phi({\bf r}_i - {\bf r}_j) = \Phi_{\rm tot.}$,
where $\Phi_{\rm tot.}$ is given by (30).
Letting then $S(E^\prime,N,V; u)$ be the equilibrium entropy of the 
system (31) we have
$$S^{(c)}(n,\Phi_{\rm tot.}) = S(E^\prime, N, V; u) - {3 \over 2} N
k \log(kT^\prime), \eqno(32)
$$
where $T^\prime = {2 \over 3k}(K_{E^\prime}/N)$ is the temperature
corresponding to the energy $E^\prime$ and the right hand side is the
configurational part of the equilibrium entropy of system (30) at
energy $E^\prime$.  (To actually compute $S(E^\prime, N. V; u)$ and
find the appropriate $u({\bf x})$ for a specified $n({\bf x})$ and
$\Phi$ we would have to use the canonical or grand canonical ensembles
and the corresponding Gibbs entropies, see section 6.)

These considerations simplify for a system of hard spheres where the
interactions do not contribute to the energy, i.e.\ $E=K$.  The phase
space domain for a specified $f$ is then just a direct product of the
configurational and momentum space regions.  The configuration space
itself is modified from what it is for a noninteracting system to
exclude all microstates $X$ such that the distance between any pair of
particles is less than $a$, the hard sphere diameter.  Eq.\ (32) now
assumes the form
$$
S_{hs}(f) = S^{(m)}(f) + {\cal S}^{(c)}_{hs}(n)
  \eqno(33)
$$
where $E$ is specified by $f$ and ${\cal S}^{(c)}_{hs}(n)$ is the
configurational part of the entropy of an equilibrium system of hard
spheres kept at a nonuniform density $n({\bf x}) = \int_{{\Bbb R}^3}
f({\bf x},{\bf v}) d{\bf v}$ by some external potential $u({\bf x})$.

The entropy function $S_{hs}(f)$ in (33) is the same as that used by
Resibois [6] (in a different form) to obtain an ${\cal H}$-theorem for
the modified Enskog equation (MEE).  It is generally believed that the
MEE, which is a heuristic extension of the BE, accurately describes
the evolution of $f_t({\bf x}, {\bf v})$ for a moderately dense hard
sphere fluid, say $na^3 \leq .1$, where $a$ is the hard sphere
diameter [5, 6].  The remarkable result, proved by Resibois in [6], is
that when $f_t({\bf x}, {\bf v})$ evolves according to the MEE then
$$
{d \over dt} S_{hs}(f_t) \geq 0   \eqno(34)
$$
with equality holding only when $f = f_{\rm eq}$.  (This is actually a
stronger statement about entropy increase than what is given by the BE
where collisions alone do not change $S_{\rm gas}(f)$ when $f$ is a
local Maxwellian.  The reason for this difference is the non-locality
of the collisions in the MEE, see [6].  We still have however
$S_{hs}(f) = S_{\rm loq. eq.}(e,{\bf u},n)$ when $f = \hat f({\bf v};
n, {\bf u}, T)$ and the hydrodynamic variables change slowly in
space.)

Resibois was driven to an expression for $S_{hs}$ equivalent to (33)
by the structure of the MEE and argued that it should have an
intrinsic significance.  Our identification of $S_{hs}(f)$ with the
Boltzmann entropy $S_B(f, E)$, i.e.\ as the $\log$ of the phase space
volume for such a nonequilibrium system, completely justifies
Resibois' intuition.  It further shows the necessity of the
modification of the original Enskog equation: as we have pointed out
repeatedly, any deterministic evolution equation arising from the
microscopic dynamics for macro-variables such as $f$ must be such that
the corresponding Boltzmann entropy satisfies an ${\cal H}$-theorem
for that equation.  The unmodified Enskog equation apparently does not
have that property [5, 6].

Let us make the statement about the increase of $S(f,E)$ a bit more
concrete.  Define 
$$
T_\Phi(n) = [\partial s^{(c)}(n,\Phi)/\partial \Phi]^{-1}
\eqno(35)
$$
where $s^{(c)}(n,\Phi)$ is the configurational entropy per unit volume 
for a system
with Hamiltonian (24) at uniform particle density $n$ and uniform
potential energy density $\Phi$.
$T_\Phi(n)$ is the inverse of the function
$\Phi(n,T)$ relating $\Phi$ to the temperature $T$ for this system.
Similarly
$$
\partial s^{(c)}(n,\Phi)/\partial n = -\mu_\Phi(n,T_\Phi)/T_\Phi =
-\mu(n,T_\Phi)/T_\Phi - {3 \over 
2} k \log T_\Phi + {\rm const.}
\eqno(36)
$$
where $\mu(n,T)$ is the chemical potential of such an equilibrium
system.  (Of course for an equilibrium system $T_\Phi = T_K = {m
\over 3} \langle ({\bf v} - {\bf u})^2 \rangle/k$ where $T_K$ is the kinetic
temperature of the system.)  Our definitions are motivated by 
(1), (11) and the structure of $H$ in (24).

Turning now to nonequilibrium systems with macro-states given by
$f({\bf x}, {\bf v})$ and $e({\bf x})$ we have, for $\Phi = \Phi({\bf x})$,
$$
S^{(c)}(n,\Phi) = \int_V d{\bf x} s^{(c)}(n({\bf x}), \Phi({\bf x}))
\eqno(37)
$$
To obtain $S^{(c)}(n, \Phi_{\rm tot.})$ we have to take the $\sup$ of
$S^{(c)}(n,\Phi)$ over all $\Phi({\bf x})$ such that
$$\int_V \Phi({\bf
x}) d{\bf x} = \Phi_{\rm tot.}.
\eqno(38)
$$
Using (35) this yields immediately, as might be expected, that the
$\sup$ is achieved when $T_{\Phi({\bf x})}(n({\bf x}))
= {\rm const.}$, which we shall call $T_{\Phi_{\rm tot.}}$ since its
value is determined by the requirement (38).

We are now able, after some manipulations, to obtain 
\vfill \eject
$$
{dS(f_t,E) \over dt} = -k \int \int(\log {f_t \over \tilde f_t})
{\partial f_t
\over \partial t} d{\bf x} d{\bf v} - \int d{\bf x} [{\mu_\Phi(n,
T_{\Phi_{\rm 
tot.}}) \over T_{\Phi_{\rm tot.}}} + {\mu_K(n,T_K) \over T_K}]{\partial
n(x,t) \over \partial t}
$$
$$
+ \int 
d{\bf x} T^{-1}_K {\partial \over
\partial t} K_{\bf u}({\bf x},t) ~~ + ~~ T^{-1}_{\Phi_{\rm tot.}} {d
\Phi_{\rm tot.} \over dt}\eqno(39)
$$
where
$$
\mu_K = -{3 \over 2} k  \log T_K({\bf
x}, t) + {\rm const.},
\eqno(40)
$$
$$T_K({\bf x},t) = {2 \over 3k} K_{\bf u}({\bf x},t)/n({\bf x},t),
$$
with
$$
K_{\bf u}({\bf x}, t) = {m \over 2} \int ({\bf v} - {\bf u}({\bf
x},t))^2 f({\bf 
x}, {\bf v}, t)d{\bf v} = K({\bf x},t) - {m \over 2} n({\bf x},t)
{\bf u}^2({\bf x},t)
$$
and where
$$
\tilde f_t = n({\bf x},t)(2 \pi m T_K)^{-3/2} \exp [-{m \over 2} ({\bf v}
- {\bf u}({\bf x},t))^2/k T_K]
\eqno(41)
$$
is a local Maxwellian with parameters $T_K, n,K$ and $\bf u$ computed
from $f_t$. The first term in (39) corresponds to changes in the entropy due
to the redistribution of velocities, while the other terms are as expected
from thermodynamics considerations.

Eq.\ (40) simplifies for a spatially uniform system for which ${\bf u}
= 0$, ${\partial n
\over \partial t} = 0$ and $T_K$ is independent of $\bf x$.  If,
furthermore, $f_t$ happens to be a Maxwellian with a temperature
$T_K$ (a case considered by Jaynes [16]), then (at that instant) 
$$
dS(f_t,E)/dt = {dK \over dt} [T_K^{-1} -
T_{\Phi_{\rm tot.}}^{-1}]. \eqno(42)
$$
where $K = E - \Phi_{\rm tot.}$ is the total kinetic energy of the
system.  Our assertion is then that $dK/dt$ has the same sign as
$(T_{\Phi_{\rm tot.}} - T_K)$, which is certainly expected even in the
absence of any deterministic equation for $f$.

\bigskip
{\bf 6.  Other Kinetic Equations}

The Boltzmann and modified Enskog equations are appropriate for
systems in which the interaction between the particles can be
represented by a succession of uncorrelated binary encounters or
collisions.  The extension of these equation to dense, non-hard sphere
fluids or to mixtures of hard spheres is not straightforward [17].
We are not aware of any results about ${\cal H}$-theorems for such equations.

{}For systems with dominant long range interactions, such as plasmas,
the time evolution of $f_t$ (where $f$ has several components) is
determined in suitable regimes of temperature and density by a Vlasov
equation combined with Boltzmann, Balescu-Lenard, or Landau collision
terms [18].  The Vlasov term describes in a mean-field way the long
range interaction between the particles.  These interactions
contribute to the energy, which is determined entirely by $f$, but not
directly to the entropy of the system, i.e.\ the entropy continues to
be given (at the level of approximation considered) by $S_{\rm
gas}(f)$ in (17).  When the short range collisions are neglected, the
smooth solutions of the Vlasov equation, like those of the Euler
equations, leave the entropy unchanged.  The inclusion of the short
range collision terms then provides an ${\cal H}$-theorem for $S_{\rm
gas}(f_t)$.  This is analogous to what happens when the viscosity
terms are added to the Euler equations to yield the Navier-Stokes
equations.

Going beyond deterministic equations, there must also be an ${\cal
H}$-theorem when the macro-variables undergo a stochastic Markovian
evolution, since, in the thermodynamic limit, the probability of a
transition to lower entropy is of much smaller order than the (order
unity) probabilities describing the Markov process.  However, we know
of no examples of such macro-variables.  (The small scale stochastic
correction to the deterministic evolution of macro-variables will of
course fail to obey an ${\cal H}$-theorem, since the probability of
these fluctuations is of the same order as the exponential of the
entropy changes.)
\bigskip \bigskip
{\bf 7. Concluding Remarks}

The entropy of a macroscopic state, defined by Eqs.\ (1) and (6), is
clearly a property of an individual macroscopic system specified by
macro-variables $M$.  We neither have nor need ensembles to observe
the time asymmetric evolution of the color profile of a glass of water
in which we dissolve a capsule of purple ink.  The appropriate choice
of $M$ for this process is clearly that corresponding to dividing the
glass into a suitable large number of little cubes and specifying the
coarse grained fraction of ink molecules in each cube as was done in
section 3.  The exact number of little cubes, as long as it is still
small compared to the number of ink molecules, will not affect $S_B$
to leading order in the number of molecules.  To this order $S_B$ will
coincide with $S_{\rm loc. eq.}$.  The evolution of $M_t$ will be
given by the solution to a diffusion equation and $S_B(M_t)$ will
satisfy the second law.

As argued in section 3, this behavior can be understood fully from
Boltzmann's microscopic interpretation of entropy.  This direct
explanatory connection between Boltzmann's entropy and the observed
behavior of individual macroscopic systems seems lacking in other
definitions of entropy in which probability distributions are a key
ingredient.  The best known of these is the Gibbs entropy,
$$
S_G(\rho) = -k \int_\Gamma \rho \log \rho dX  \eqno(43) $$
where $\rho(X)$ is some given probability  (ensemble) density.
Clearly if $\rho =  \rho_M$
$$
 \rho_M(X) = \cases{ |\Gamma_M|^{-1}, & if $X \in \Gamma_M$\cr
0, & otherwise.\cr}  \eqno(44)
$$
then
$$
S_G(\rho_M) = k \log |\Gamma_M| = S_B(M) \eqno(45)
$$

{}For a system specified by the thermodynamic variables $M(X) =
(E,N,V)$, $\rho_M$ corresponds to the microcanonical ensemble.  Using
the equivalence of equilibrium ensembles for macroscopic systems, Eq.\
(45) holds also for the canonical and grandcanonical ensembles, in the
thermodynamic limit.  In fact $S_G$ is of paramount importance both in
the mathematical foundations and practical applications of equilibrium
statistical mechanics.  On the other hand, as is very well known,
$S_G$, unlike $S_B$, does not change in time for an isolated system
evolving under Hamiltonian dynamics [$S_G(t) = S_G(\rho_t)$, where
$\rho_t(X)$ is $\rho_0(X_{-t})$].  It is therefore inappropriate, we
believe, to use $S_G$ or quantities like it in ``derivations'' of the
second law without explicitly considering typical behavior.  Attempts
to remedy this through the use of coarse grained ensembles may be
useful mathematically, but conceptually they are just variations on
the Boltzmann entropy [19].

The Boltzmann entropy itself is, as indicated by the use of
macro-variables for its very formulation, meaningful only for
macroscopic systems.  {}For such systems one can speak of the behavior
of the macro-variables $M$ as arising from the evolution of a typical
microstate in $\Gamma_{M_0}$.  It might still be true that a system
containing just a few particles exhibits ergodicity, mixing, positive
Lyapunov exponents, etc.; this is true e.g.\ for a particle moving
among fixed convex scatterers on a torus (Sinai billiard).  But the
physical system of one such particle will not exhibit any time
asymmetric behavior, corresponding to the diffusion of the purple ink
in the glass of water.  Unlike the glass of inky water or a very large
number of particles moving among such scatterers (see discussion at
end of sec.\ 4) a film of the particle's motion run backwards will
look the same as one run forward.

The situation is different when one considers open systems, e.g.\
systems in contact with thermal reservoirs [19, 20].  The time
evolution of the microstate $X$ of such a system is then no longer
given by a Hamiltonian since the system is not isolated and
$S_G(\rho_t)$ need no longer be constant in time.  It is in fact
reasonable in some cases to treat $X$ as a random variable evolving
via a stochastic Markovian dynamics.  It is then easy to show that
when the Markov process has a stationary density ${\bar \rho}(X)$ then
the relative entropy
$$
S_G(\rho_t|\bar \rho) = - k \int_\Gamma \rho_t \log(\rho_t / \bar
\rho)dX  \eqno(46) 
$$
increases monotonically in time.

A different situation, of current interest, in which $S_G(\rho)$ is
not constant is that of a closed system evolving under a deterministic
non-Hamiltonian thermostated dynamics [19, 20].  Starting with an
initial density $\rho(X,0)$, uniform (or absolutely continuous) with
respect to the appropriate Lebesgue measure, the dynamics leads to
$\rho(X,t) \to \bar
\rho$, as $t \to \infty$, with $\bar \rho$  singular with respect to 
Lebesgue measure, ($\bar \rho$ is generally an SRB measure), and 
$S_G(\rho) \to -\infty$ decreasing with time in such a way that,
$$
{d \over dt} S_G(\rho) \to -\sigma, \eqno(47)
$$
with $\sigma > 0$.
This $\sigma$ is frequently interpreted as an ``entropy production''
and indeed has, in some cases, a form similar to the hydrodynamic
entropy production in a stationary open system in which there are
currents.  The connection between the steady states of systems
evolving under the invented thermostated dynamics and those obtained
from more realistic models for which the stationary $\bar \rho$ is not
singular, is still not entirely clear.  The same is true for the
correct identification of entropy in such systems [17--23].

It should be noted that the word entropy is used very widely in
contexts other than that of macroscopic physical systems.  There is
the Shannon information entropy--- which is formally similar to $S_G$,
but is designed for situations which have apparently nothing to do
with thermodynamics---the Kolmogorov-Sinai and topological entropies of
dynamical systems, etc..  These entropies are clearly very useful and
clearly different from the Clausius and Boltzmann entropies.
Surprisingly, there are frequently some unexpected deep connections
between these different entropies which are very interesting [24, 25, 26].

Let us finally discuss the choice of appropriate macro-variables $M$
in terms of which to describe a particular nonequilibrium system of
interest in some microstate $X$ which is far from LTE.  We do not want
to choose too coarse a description: one for which $X$ is not typical
of $\Gamma_{M(X)}$.  This occurred in the example of Jaynes discussed
in section 5; the problem could be remedied there by introducing more
refined macrostates, given not just by $f$ but by $f$ and $E$.  We
also do not want to use a description more detailed than is relevant
to macroscopic behavior.

What we are after is a useful minimal description via macro-variables
$M$, adapted to the situation (corresponding to a microstate $X$)
under consideration.  This is always achieved when $M$ is such that
(1) $M_{X_t}$ typically obeys an autonomous deterministic evolution
law, such as those corresponding to hydrodynamics, the BE, or the MEE,
and (2) $X$ is typical in this sense of $\Gamma_{M(X)}$ .
\bigskip
\noindent {\bf Acknowledgments}:  We want to thank Raffaello Esposito, 
Giovanni Gallavotti, Michael Kiessling, Christian Maes, David Ruelle
and Roderich Tumulka for useful discussions.  J.L. also want to thank
Harry Swinney and Constantino Tsallis for their hospitality at the
International Workshop on Anomalous Distributions, Nonlinear Dynamics
and Nonextensivity, Santa Fe, November, 2002, which prompted this
article.  Research supported by NSF Grant DMS 01-279-26, and AFOSR
Grant AF 49620-01-1-0154.

\bigskip

\noindent {\bf References}

\item {[1]}  c.f.\ H. B. Callen, {\it Thermodynamics, and an
Introduction to Thermostatistics}, Wiley (1960).

\item {[2]}  S. R. de Groot and P. Mazur, {\it Nonequilibrium
Thermodynamics}, North-Holland (1959).

\item {[3]}  For a historical survey see S. G. Brush, {\it The 
Kind of Motion We Call Heat}, North-Holland (1976);  M. Klein, The
Development of Boltzmann's Statistical Ideas, in {\it The Boltzmann
Equation}, ed. E. G. D. Cohen, W. Thirring, Acta Physica Austriaca
suppl. X, Vienna, p. 53 (1973).

\item {[4a]}  For a collection of basic papers from the second half of
the nineteenth century on the subject (in English translation), see
S. G. Brush, {\it Kinetic Theory}, {\bf 1, 2}, Pergamon, Elmsford,
N.Y. (1965-1966).  The apparently little known 1874 article by 
W. Thomson, (Lord Kelvin), The Kinetic Theory of
the Dissipation of Energy, Proc. Royal Soc. Edinburgh {\bf 8},
325-328, (1874), reprinted there, 
is highly recommended.

\item {[4b]}  L. Boltzmann, {\it Vorlesungen {\"u}ber Gastheorie}, 2
vols. Leipzig:  Barth, 1896, 1898.  This book has been translated into
English by S. G. Brush, {\it Lectures on Gas Theory}, (Cambridge
University Press, London, 1964, reprinted Dover, 1995).

\item {[5]}  C. Cercignani, {\it The Boltzmann Equation and Its
Applications}, Springer-Verlag, (1988).

\item {[6]}  P. Resibois, H-Theorem for the (Modified) Nonlinear
Enskog Equation, {\it J. Stat. Phys}, {\bf 19}, 593 (1978).

\item {[7a]}  J. L. Lebowitz, Boltzmann's
Entropy and Time's Arrow, {\it Physics Today}, {\bf 46}, 32--38, 1993;
Microscopic Reversibility and Macroscopic Behavior: Physical
Explanations and Mathematical Derivations, in {\it 25 Years of
Non-Equilibrium Statistical Mechanics}, Proceedings, Sitges
Conference, Barcelona, Spain, 1994, in Lecture Notes in Physics,
J.J. Brey, J. Marro, J.M. Rub{\'i} and M. San Miguel (eds.), Springer-Verlag,
1995; Microscopic Origins of Irreversible Macroscopic Behavior, {\it
Physica A}, {\bf 263}, 516--527, 1999.

\item {[7b]}  S. Goldstein, 
 Boltzmann's Approach to Statistical Mechanics, in Chance in Physics:
{}Foundations and Perspectives, edited by Jean Bricmont et al., Lecture
Notes in Physics 574, (Springer-Verlag, 2001), cond-mat/0105242.

\item {[8]}  R. Clausius, The Nature of the Motion we Call Heat, {\it
Ann. Phys.} [2]. {\bf 125}. 353 (1865), reprinted in [4a].

\item {[9]} L. Onsager , Thermodynamics and Some Molecular Aspects of
Biology, in {\it The Neurosciences}, A Study Program,
eds. G. C. Quarton et al. (Rockefeller University Press, New York,
1967) pp. 75-79.

\item {[10]}  H. Spohn, {\it Large Scale Dynamics of Interacting
Particle Systems}, Springer-Verlag (1991).

\item {[11]}  L. Landau and S. Lifshitz, {\it Fluid Mechanics},
Butterworth-Heinemann (1987).

\item {[12]}  G. Gallavotti, {\it Foundations of Fluid Dynamics},
Springer-Verlag (2002).

\item {[13]}  {\it Albert Einstein, Philosopher-Scientist}.
Autobiographical Notes, p. 43, in The Library of Living Philosophers
(1949) (sixth printing 1995), P. A. Schlipp, editor.

\item {[14]}  O. Lanford, Time Evolution of Large Classical Systems,
J. Moder, ed., LND {\bf 38}, Springer-Verlag (1975);  {\it Physica}, A{\bf
106}, 70 (1981). 

\item {[15]}  E. Carlen and M. Carvalho, Entropy Production Estimates
for Boltzmann Equation with Physically Realistic Collision Kernels,
{\it J. Stat. Phys.} {\bf 74}, 743-782, (1994); L. Desvillettes and
C. Villani, On the Trend to Global Equilibrium for Spatially
Inhomogeneous Kinetic Systems: The Boltzmann Equation, preprint.

\item {[16]}  E. T. Jaynes, Violation of Boltzmann's H Theorem in Real
Gases, {\it Phys. Rev. A}, {\bf 4}, 747-750 (1971). 

\item {[17}]  J. A. McLennan, {\it Introduction to Non-Equilibrium
Statistical Mechanics}, Prentice Hall (1989).

\item {[18]}  R. Balescu, {\it Transport Processes in Plasmas},
North-Holland (1988).

\item {[19]}  C. Maes and K. Netocny, Time-reversal and Entropy, {\it
J. Stat. Phys.} {\bf 110}, 269-310
(2003).

\item {[20]}  See e.g.\ F.~Bonetto, J.~L. Lebowitz and L.~Rey-Bellet,
{F}ourier's law: a challenge to theorists, in 
{\it Mathematical
Physics 2000\/}, pages 128--150, London, 2000. Imperial College Press,
A.~ Fokas, A.~Grigoryan, T.~Kibble and B.~Zegarlinski (eds.)

\item {[21]}  G. Gallavotti and E. G. D. Cohen, Dynamical Ensembles in
Nonequilibrium Statistical Mechanics, {\it Phys. Rev. Lett.} {\bf 95},
74, 2694 (1995); G. Gallavotti, Nonequilibrium Thermodynamics,
preprint  (2002).

\item {[22]}  D. Ruelle, Smooth Dynamics and New Theoretical Ideas in
Nonequilibrium Statistical Mechanics, {\it J. Stat. Phys.} {\bf 95},
393--468 (1999); Extending the Definition of Entropy to Nonequilibrium
Steady States, Proc. Natl. Acad. Sci. USA, 3054--3058, {\bf 100}
(2003). math phys archive 03-98..

\item {[23]}  N. Chernov and J. L. Lebowitz, "Stationary
Nonequilibrium States in Boundary Driven amiltonian Systems:  Shear
Flow, {\it Journal of Statistical Physics}, {\bf 86}, 953--990, 1997.
Los Alamos cond-mat/9703097.  Texas 97--113.

\item {[24]}  Ya. G. Sinai {\it Introduction to Ergodic Theory},
Lecture 14, Princeton University Press (1976).

\item {[25]}  G. Gallavotti, {\it Statistical Mechancis:  A Short
Treatise}, Springer-Verlag (1999).

\item {[26]}  {\it Dynamical Systems, Ergodic Theory and
Applications}, edited by Ya. Sinai, Springer-Verlag (2000).

\end